# Mining Positive and Negative Association Rules Using CoherentApproach

RAKESH DUGGIRALA[1], P.NARAYANA[2]

[1] M.Tech(CSE), **Gudlavalleru Engineering College, Gudlavalleru**
[2] **Associate professor (C.S.E), Gudlavalleru Engineering College, Gudlavalleru**

**ABSTRACT:**

**Abstract**—In the data mining field, association rules are discovered having domain knowledge specified as a minimum support threshold. The accuracy in setting up this threshold directly influences the number and the quality of association rules discovered. Typically, before association rules are mined, a user needs to determine a support threshold in order to obtain only the frequent item sets. Having users to determine a support threshold attracts a number of issues. We propose an association rule mining framework that does not require a pre-set support threshold.

Often, the number of association rules, even though large in number, misses some interesting rules and the rules' quality necessitates further analysis. As a result, decision making using these rules could lead to risky actions. We propose a framework to discover domain knowledge report as coherent rules. Coherent rules are discovered based on the properties of propositional logic, and therefore, requires no background knowledge to generate them. From the coherent rules discovered, association rules can be derived objectively and directly without knowing the level of minimum support threshold required. We provide analysis of the rules compare to those discovered via the apriori. The framework is developed based on implication of propositional logic via Negative and positive association algorithm. The experiments show that our approach is able to identify meaningful association rules within an acceptable execution time. This framework develop a new algorithm based on coherent rules so that users can mine the items without domain knowledge and it can mine the items efficiently when compared to association rules.

**I INTRODUCTION**

The process of extracting information from large databases is termed as data mining. Different algorithms are used to extract the data, by using association rules frequent items are discovered from database by specifying minimum support threshold value. The items mined require domain knowledge [3] and statistical methods to specify that threshold value. If the threshold value is very high rare items may be missing, if it is low there may be inconsistency in the items retrieved. Some interesting rules are missing for future analysis which leads to errors in decision making, for mining rare items [12] they are grouped into arbitrary items becomes frequent items this is called rare items problem defined by Manila[13] according to Liu et.al. and another method is to split the data set into two or several blocks according to the frequency and mine each block using minimum support threshold although some association rules involving both frequent and rare items across different blocks is lost.

Later frequent items are mined using multiple minimum threshold called minimum item support (MISs) by Liu et.al [12],later items are mined using minimum relative support and minimum confidence even though they are not properly correlated [2] then by using lift, leverage the minimum threshold the items are mined and they are not asymmetrical so to overcome this an approach of coherent rules using implications is used to mine the items through which the associations rules are discovered inherently, using some standard logic tables called implications. The frequent items are extracted which don☐t require specification of minimum support threshold and the items can be mined without domain knowledge of user. The association rules can be derived inherently by using these coherent rules.

Association rule mining, introduced considered as one of the most important tasks in Knowledge Discovery in Databases . Among sets of items in transaction databases, it aims at discovering implicative tendencies that can be valuable information for the decision-maker. An association rule is defined as the implication X→Y, described by two interestingness measures support and confidence, where X and Y are the sets of items and X AND Y = φ . Apriori [2] is the first algorithm proposed in the association rule mining field and many other algorithms were derived from it. It is very well known that mining algorithms can discover a prohibitive amount of association rules; Starting from a database, it proposes to extract all association rules satisfying minimum thresholds of support and confidence. For instance, thousands of rules are extracted from a database of several dozens of attributes and several hundreds of transactions.

Discovering coherent rules resolves the many difficulties in mining associations that require a preset minimum support threshold. Apart from solving the issues of a support threshold, the coherent rules found can also be reasoned as logical implications due to the mapping to the truth table values of logical equivalence. In contrast, classic association rules cannot be reasoned as logical implications due to their lack of this logic property.

The popularity and importance of data mining has its roots in two causes: the ever-increasing volume of data and computation power. The amount of information in the world doubles every twenty months [FrPiMa92]. Business activities, for example, continue to produce an increasing stream of data (such as point-of-sales transactions) which is stored in larger and cheaper data storage. In the meantime, the computational power available continues to increase. Gordon Moore, co-founder of the Intel corporation, points out that the number of transistors on a chip doubles approximately every two years [In05a], and that this trend has continued for more than half a century [In05b]. The consequence of the increasing volume of data and computational power is an opportunity to create data mining applications based on state-of-art theories and algorithms to discover interesting knowledge from large volumes of data.

## 2. LITERATURE SURVEY

More recently, it is accepted that infrequent rules are also important because it represents knowledge not found in frequent rules, and these infrequent rules are often interesting [1] In addition to missing infrequent item rules, the traditional algorithm such as apriori [3] also does not report the existence of negative associations. Association among infrequent items and negative associations have been relatively ignored by association mining algorithm mainly due to the problem of large search space and the explosion of total number of association rules reported. Some of these rules may in fact are noise in the data. There are some attempts to find infrequent association such as that of [9]. This work proposed a generalise association using correlation. Correlation is measured by chi-square. However, at small expected values, the measure of chi-square has limitation of measuring the association accurately and, hence, results may be inaccurate. In addition, the authors' algorithm relies on a modified support hence, is not really suitable to find infrequent rules except the ones that are above a threshold. finds independent rules measured by interest (leverage) and below a minimum support threshold. Authors in [1] also use [1] measure, which is derived from correlation, and necessitates a minimum confidence threshold. Mining below a minimum support threshold is similar to having a maximum support threshold. In addition, measure used in [1] will inherit the drawbacks of a correlation measure in [1] filters uninteresting rules using leverage as a measure. [1],[2] finds rules using measure such as leverage or lift; these can be performed without other thresholds in place. Since rules found are independent from a minimum support threshold, theoretically it could find all infrequent rules. Rules found using leverage however measures co-occurrence but not the real implication [5].

The research by authors Liu, Hsu and Ma [4], Lin, Tseng and Su [7], Yun et al. [8] and Koh, Rountree and O'Keefe [5] has been important in establishing a minimum item support threshold with finer granularity although different criteria were injected for identifying minimum item support values. The common aim, however, was to offset heuristics when setting up a minimum support threshold. In all these approaches, we see that state-of-art association rule mining has drifted from the original idea of mining frequent patterns alone to considering other patterns as well.

Omiecinski has another approach finding rules with high confidence values. This author proposed measures of interestingness which are called all-confidence and bond, which were shown to satisfy the anti-monotone property.

All-confidence means that all association rules produced from an item set would have the confidence of at least the all-confidence value whereas bond is a symmetrical measure and a special case within all-confidence. By setting a minimum threshold on bond, one can discover association rules where their all-confidence values are at least the bond values.

In much of association rule mining, data sets revolve around transaction records. During the mining, the content of transaction record is observed. As highlighted in the motivation section, observing the absence of items from a transaction record will produce a more complete result. The absence of an item is described in the following example. Assume there are three items in a dataset, items A, B and C. A transaction contains only item A. Items B and C are said to be absent from this transaction.

Cornelis et al. [CoYaZhCh06] avoided using leverage to search for negative association rules as leverage does not inherit anti-monotone property. Instead, they developed a procedure to generate item sets for both positive and negative association rules using a pre-set minimum support threshold.

## 3. PROPOSED FRAMEWORK

An implication that meets both the two contrapositives is an implication of equivalence. This is a more stringent implication and is a special case in material implication. We are interested in an association rule framework that maps to the logical equivalences to find non-trivial association rules. Based on the logic property of equivalence, we can consider both the presence and absence of item sets in a set of transaction records, without this process requiring a minimum support threshold to identify association rules. Apart from not requiring a minimum support threshold, an implication of logical equivalence can avoid contradictions in reasoning with found rules.

**Association Rules and Supports**

| Association Rule | Support |
|---|---|
| $X \Rightarrow Y$ | $S(X, Y)$ |
| $X \Rightarrow \neg Y$ | $S(X, \neg Y)$ |
| $\neg X \Rightarrow Y$ | $S(\neg X, Y)$ |
| $\neg X \Rightarrow \neg Y$ | $S(\neg X, \neg Y)$ |

We give the name pseudo-implication to association rules that are mapped to implications based on comparison between supports. By pseudo-implication, we mean that the implication approximates a real implication (according to propositional logic). It is not a real implication because there are fundamental differences. Pseudo-implication is judged true or false based on a comparison of supports, which has a range of integer values. In contrast, an implication is based on binary values. The former depends on the frequencies of co-occurrences between item sets (supports) in a dataset, whereas the latter does not and is based on truth values.

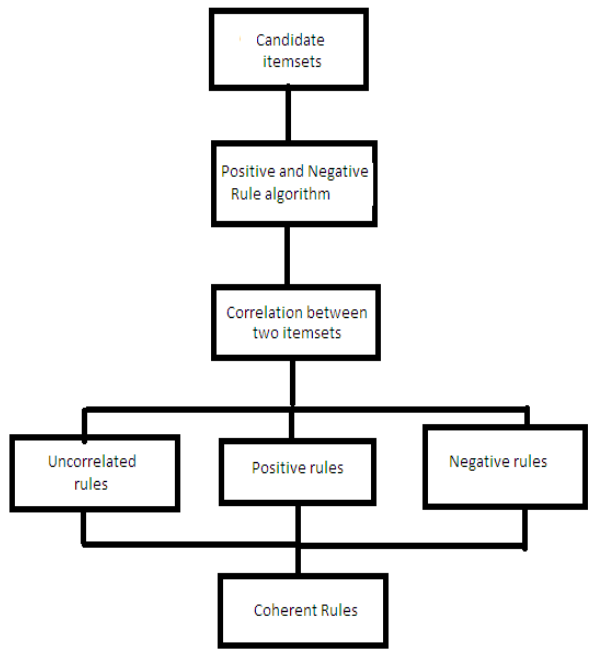

The algorithm generates all positive and negative association rules that have a strong correlation. If no rule is found, either positive or negative, the correlation threshold is automatically lowered to ease the constraint on the strength of the correlation and the process is redone. Figure 1 gives the detailed pseudo-code for our algorithm.

Initially both sets of negative and positive association rules are set to empty (line 1). After generating all the frequent 1-itemsets (line 2) we iterate to generate all frequent k-itemsets, stored in Fk (line 8). Fk is verified from a set of candidate Ck computed in line 4. The iteration from line 2 stops when no longer frequent itemsets are possible. Unlike the join made in the traditional Apriori algorithm, to generate candidates at level k, instead of joining frequent (k -1)-itemsets, we join the frequent itemsets at level k -1 with the frequent 1-itemsets (line 4). This is because we want to extend the set of candidate itemsets and have the possibility to analyze the correlation of more item combinations. The rational will be explained later. Every candidate itemset generated this way is on one hand tested for support (line 7), and on the other hand used to analyze possible correlations even if its support is below the minimum support (loop from line 9 to 22). Correlations for all possible pair combinations for each candidate itemset are computed. For an itemset i and a pair combination (X; Y ) such that i = X U Y , the correlation coefficient is calculated (line 10). If the correlation is positive and strong enough, a positive association rule of the type X □ Y is generated, if the supp(X UY ) is above the minimum support threshold and the confidence of the rule is strong.

**Algorithm: Negative association rules .**
Input:
D:Transactional database,
ms: minimum support,
mc: minimum confidence
Output: Positive and Negative Association Rules
Method**:**
(1) L1=frequent-1-positive-itemsets(D)
(2) N1=frequent-1-Negative-itemsets(D)
// complement frequent-1-positive-itemsets(D)
(3) L=L1 U N1;
(4) for (k=2; Lk-1 ≠ Ø; k++)
(5) {
(6) **//** Generating Ck
(7) for each l1,l2 ε Lk-1
(8) If(l1[1]=l2[1]^............l1[k-2]=l2[k-2]^l1[k-1]<l2[k-1])
(9) Ck=Ck U {l1 [1].......l1 [k-2], l1[k-1], l2[k-1]}
(10) end if
(11) end for
(12) **//** Pruning using Apriori property
(13) for each (k-1)- subsets s of c in Ck
(14) If s is not a member of Lk-1
(15) Ck=Ck – {c}
(16) end if
(17) end for
(18) PCk= Ck;
(19) for each c in PCK
(20) NCk={ck
1/ ck
1 is obtained by replacing a
literal of c in PCk by its negation}
(21) //Pruning using Support Count
(22) Scan the database and find support for all c in PCk
(23) Lk=candidates in PCk that pass support threshold
(24) Find support for all ck
1 in NCk from supports
of members of PCK and Lk-1
(25) Nk= candidates in NCk that pass support threshold
(26) L= Lk U Nk
(27) }

Line 1 generates positive-frequent-1- itemsets
 Line 2 generates negative-frequent-1- itemsets by complementing 1-itemsets
obtained in Line 1
 Line 8 and 9 generates candidate itemsets Ck using Apriori algorithm
 Line 13-15 pruning candidate itemsets in Ck using Apriori property
 Lines 18, after pruning, the remaining elements are treated as valid candidates and is denoted by PCk.
 Line 19-20, for each literal of this valid candidate, replace the literal with the
corresponding negated literal, creates a new negative rule and denoted by NCk.
Each valid candidate with n number of literals in the antecedent will generate n new negative itemsets. For example a 3- itemset ABC will give 3 negative items ¬ABC, A¬BC, and AB¬C.
 Line 22-23, prune all items in PCk using support count and add to Lk, set of
frequent k-itemsets
 Line 24, find support count of all items in NCk using PCk and Lk-1.
 support(¬A) = 1- suuport(A)
 support(AU¬B) = support(A)-support(AU B)

- support(¬A∪B) = support(B)-support(A∪B)
- support(¬A∪¬B) = 1-support(A)- support(B) + support(A∪B)
- Line 25, Nk is the set of all elements whose support ≥ minsupp.
- The generation of positive rules continues without disruption and the rich but valuable negative rules are produced as by-products of the Apriori process.

## 4. EXPERIMENTAL RESULTS

ARM RULES:
=======

Minimum support: 0.6 (61 instances)
Minimum metric <confidence>: 0.9
Number of cycles performed: 8

Generated sets of large itemsets:

Best rules found:

 1. venomous=false tail=true 71 ==> backbone=true 71    conf:(1)
 2. aquatic=false 65 ==> fins=false 65    conf:(1)
 3. aquatic=false breathes=true 64 ==> fins=false 64    conf:(1)
 4. backbone=true venomous=false fins=false 63 ==> breathes=true 63    conf:(1)
 5. toothed=true 61 ==> feathers=false 61    conf:(1)
 6. toothed=true 61 ==> backbone=true 61    conf:(1)
 7. toothed=true backbone=true 61 ==> feathers=false 61    conf:(1)
 8. feathers=false toothed=true 61 ==> backbone=true 61    conf:(1)
 9. toothed=true 61 ==> feathers=false backbone=true 61    conf:(1)
10. aquatic=false venomous=false 61 ==> fins=false 61    conf:(1)
11. venomous=false tail=true domestic=false 61 ==> backbone=true 61    conf:(1)
12. tail=true 75 ==> backbone=true 74    conf:(0.99)
13. backbone=true fins=false 66 ==> breathes=true 65    conf:(0.98)
14. aquatic=false 65 ==> breathes=true 64    conf:(0.98)
15. aquatic=false fins=false 65 ==> breathes=true 64    conf:(0.98)
16. aquatic=false 65 ==> breathes=true fins=false 64    conf:(0.98)
17. tail=true domestic=false 65 ==> backbone=true 64    conf:(0.98)
18. backbone=true breathes=true 69 ==> venomous=false 67    conf:(0.97)
19. backbone=true breathes=true fins=false 65 ==> venomous=false 63    conf:(0.97)
20. feathers=false backbone=true 63 ==> airborne=false 61    conf:(0.97)
21. feathers=false backbone=true 63 ==> toothed=true 61    conf:(0.97)
22. backbone=true tail=true 74 ==> venomous=false 71    conf:(0.96)
23. backbone=true fins=false 66 ==> venomous=false 63    conf:(0.95)
24. backbone=true fins=false 66 ==> breathes=true venomous=false 63    conf:(0.95)
25. backbone=true tail=true domestic=false 64 ==> venomous=false 61    conf:(0.95)
26. backbone=true 83 ==> venomous=false 79    conf:(0.95)
27. breathes=true 80 ==> fins=false 76    conf:(0.95)
28. airborne=false 77 ==> feathers=false 73    conf:(0.95)
29. tail=true 75 ==> venomous=false 71    conf:(0.95)
30. tail=true 75 ==> backbone=true venomous=false 71    conf:(0.95)
31. breathes=true venomous=false 75 ==> fins=false 71    conf:(0.95)
32. airborne=false venomous=false 71 ==> feathers=false 67    conf:(0.94)
33. backbone=true domestic=false 71 ==> venomous=false 67    conf:(0.94)
34. backbone=true breathes=true 69 ==> fins=false 65    conf:(0.94)
35. airborne=false domestic=false 68 ==> feathers=false 64    conf:(0.94)
36. breathes=true domestic=false 68 ==> venomous=false 64    conf:(0.94)
37. breathes=true domestic=false 68 ==> fins=false 64    conf:(0.94)
38. backbone=true breathes=true venomous=false 67 ==> fins=false 63    conf:(0.94)
39. aquatic=false 65 ==> venomous=false 61    conf:(0.94)
40. airborne=false backbone=true 65 ==> feathers=false 61    conf:(0.94)
41. airborne=false backbone=true 65 ==> venomous=false 61    conf:(0.94)
42. aquatic=false fins=false 65 ==> venomous=false 61    conf:(0.94)
43. aquatic=false 65 ==> venomous=false fins=false 61    conf:(0.94)
44. tail=true domestic=false 65 ==> venomous=false 61    conf:(0.94)
45. tail=true domestic=false 65 ==> backbone=true venomous=false 61    conf:(0.94)
46. breathes=true 80 ==> venomous=false 75    conf:(0.94)
47. breathes=true fins=false 76 ==> venomous=false 71    conf:(0.93)
48. airborne=false 77 ==> venomous=false 71    conf:(0.92)
49. venomous=false fins=false 77 ==> breathes=true 71    conf:(0.92)
50. domestic=false 88 ==> venomous=false 81    conf:(0.92)
51. feathers=false venomous=false 73 ==> airborne=false 67    conf:(0.92)
52. feathers=false airborne=false 73 ==> venomous=false 67    conf:(0.92)
53. fins=false 84 ==> venomous=false 77    conf:(0.92)
54. fins=false domestic=false 72 ==> venomous=false 66    conf:(0.92)
55. backbone=true breathes=true 69 ==> venomous=false fins=false 63    conf:(0.91)
56. airborne=false domestic=false 68 ==> venomous=false 62    conf:(0.91)
57. backbone=true venomous=false domestic=false 67 ==> tail=true 61    conf:(0.91)

=== Evaluation ===

Elapsed time: 0.05s

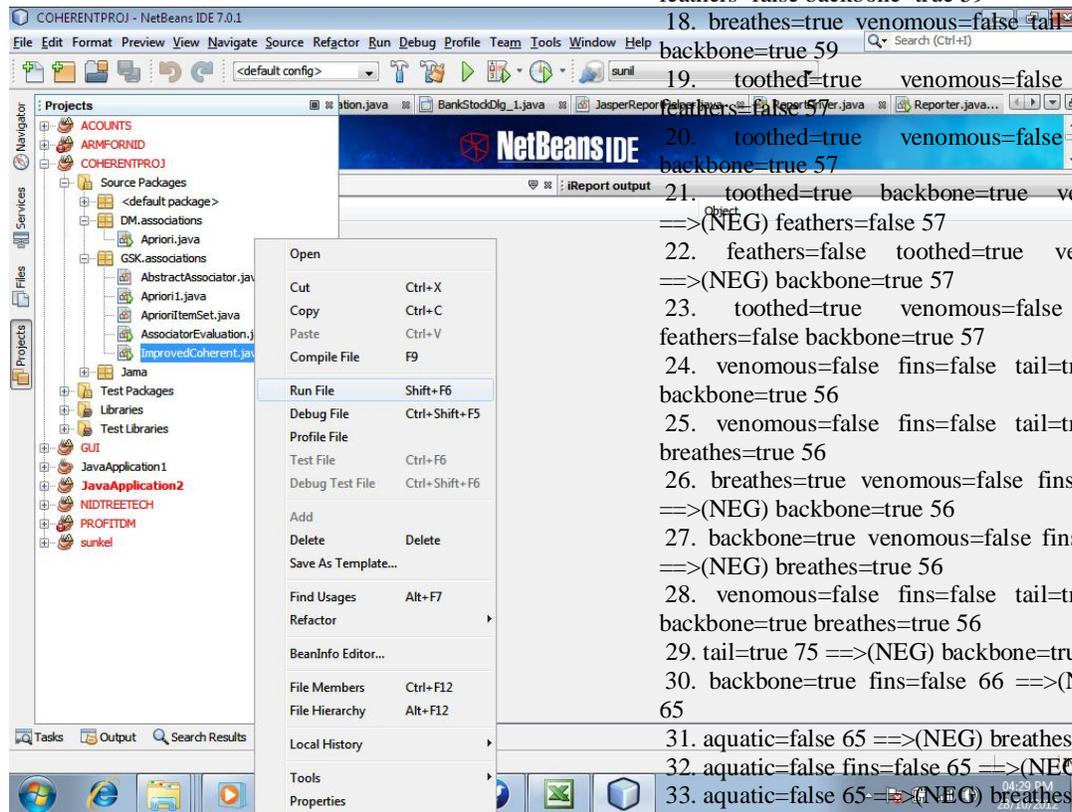

ARM RULES:
=======

 1. venomous=false tail=true 71 ==>(NEG) backbone=true 71
 2. aquatic=false 65 ==>(NEG) fins=false 65
 3. aquatic=false breathes=true 64 ==>(NEG) fins=false 64
 4. backbone=true venomous=false fins=false 63 ==>(NEG) breathes=true 63
 5. toothed=true 61 ==>(NEG) feathers=false 61
 6. toothed=true 61 ==>(NEG) backbone=true 61
 7. toothed=true backbone=true 61 ==>(NEG) feathers=false 61
 8. feathers=false toothed=true 61 ==>(NEG) backbone=true 61
 9. toothed=true 61 ==>(NEG) feathers=false backbone=true 61
 10. aquatic=false venomous=false 61 ==>(NEG) fins=false 61
 11. venomous=false tail=true domestic=false 61 ==>(NEG) backbone=true 61
 12. aquatic=false breathes=true venomous=false 60 ==>(NEG) fins=false 60
 13. airborne=false toothed=true 59 ==>(NEG) feathers=false 59
 14. airborne=false toothed=true 59 ==>(NEG) backbone=true 59
 15. airborne=false toothed=true backbone=true 59 ==>(NEG) feathers=false 59
 16. feathers=false airborne=false toothed=true 59 ==>(NEG) backbone=true 59
 17. airborne=false toothed=true 59 ==>(NEG) feathers=false backbone=true 59
 18. breathes=true venomous=false tail=true 59 ==>(NEG) backbone=true 59
 19. toothed=true venomous=false 57 ==>(NEG) feathers=false 57
 20. toothed=true venomous=false 57 ==>(NEG) backbone=true 57
 21. toothed=true backbone=true venomous=false 57 ==>(NEG) feathers=false 57
 22. feathers=false toothed=true venomous=false 57 ==>(NEG) backbone=true 57
 23. toothed=true venomous=false 57 ==>(NEG) feathers=false backbone=true 57
 24. venomous=false fins=false tail=true 56 ==>(NEG) backbone=true 56
 25. venomous=false fins=false tail=true 56 ==>(NEG) breathes=true 56
 26. breathes=true venomous=false fins=false tail=true 56 ==>(NEG) backbone=true 56
 27. backbone=true venomous=false fins=false tail=true 56 ==>(NEG) breathes=true 56
 28. venomous=false fins=false tail=true 56 ==>(NEG) backbone=true breathes=true 56
 29. tail=true 75 ==>(NEG) backbone=true 74
 30. backbone=true fins=false 66 ==>(NEG) breathes=true 65
 31. aquatic=false 65 ==>(NEG) breathes=true 64
 32. aquatic=false fins=false 65 ==>(NEG) breathes=true 64
 33. aquatic=false 65 ==>(NEG) breathes=true fins=false 64
 34. tail=true domestic=false 65 ==>(NEG) backbone=true 64
 35. aquatic=false venomous=false 61 ==>(NEG) breathes=true 60
 36. breathes=true tail=true 61 ==>(NEG) backbone=true 60
 37. aquatic=false venomous=false fins=false 61 ==>(NEG) breathes=true 60
 38. aquatic=false venomous=false 61 ==>(NEG) breathes=true fins=false 60
 39. backbone=true breathes=true tail=true 60 ==>(NEG) venomous=false 59
 40. eggs=true 59 ==>(NEG) milk=false 58
 41. fins=false tail=true 59 ==>(NEG) backbone=true 58
 42. fins=false tail=true 59 ==>(NEG) breathes=true 58
 43. breathes=true fins=false tail=true 58 ==>(NEG) backbone=true 57
 44. backbone=true fins=false tail=true 58 ==>(NEG) breathes=true 57
 45. airborne=false tail=true 57 ==>(NEG) backbone=true 56
 46. backbone=true breathes=true fins=false tail=true 57 ==>(NEG) venomous=false 56
 47. backbone=true breathes=true 69 ==>(NEG) venomous=false 67
 48. backbone=true breathes=true fins=false 65 ==>(NEG) venomous=false 63
 49. feathers=false backbone=true 63 ==>(NEG) airborne=false 61
 50. feathers=false backbone=true 63 ==>(NEG) toothed=true 61
 51. toothed=true 61 ==>(NEG) airborne=false 59

52. feathers=false toothed=true 61 ==>(NEG) airborne=false 59
53. toothed=true 61 ==>(NEG) feathers=false airborne=false 59
54. toothed=true backbone=true 61 ==>(NEG) airborne=false 59
55. toothed=true 61 ==>(NEG) airborne=false backbone=true 59
56. breathes=true tail=true 61 ==>(NEG) venomous=false 59
57. feathers=false toothed=true backbone=true 61 ==>(NEG) airborne=false 59
58. feathers=false airborne=false backbone=true 61 ==>(NEG) toothed=true 59
59. toothed=true backbone=true 61 ==>(NEG) feathers=false airborne=false 59
60. feathers=false toothed=true 61 ==>(NEG) airborne=false backbone=true 59
61. toothed=true 61 ==>(NEG) feathers=false airborne=false backbone=true 59
62. breathes=true tail=true 61 ==>(NEG) backbone=true venomous=false 59
63. milk=false 60 ==>(NEG) eggs=true 58
64. feathers=false backbone=true venomous=false 59 ==>(NEG) airborne=false 57
65. feathers=false backbone=true venomous=false 59 ==>(NEG) toothed=true 57
66. fins=false tail=true 59 ==>(NEG) backbone=true breathes=true 57
67. hair=false 58 ==>(NEG) milk=false 56
68. backbone=true breathes=true domestic=false 58 ==>(NEG) venomous=false 56
69. backbone=true fins=false tail=true 58 ==>(NEG) venomous=false 56
70. breathes=true fins=false tail=true 58 ==>(NEG) venomous=false 56
71. breathes=true fins=false tail=true 58 ==>(NEG) backbone=true venomous=false 56
72. backbone=true fins=false tail=true 58 ==>(NEG) breathes=true venomous=false 56
73. backbone=true tail=true 74 ==>(NEG) venomous=false 71
74. backbone=true fins=false 66 ==>(NEG) venomous=false 63
75. backbone=true fins=false 66 ==>(NEG) breathes=true venomous=false 63
76. backbone=true tail=true domestic=false 64 ==>(NEG) venomous=false 61
77. backbone=true 83 ==>(NEG) venomous=false 79
78. breathes=true tail=true 61 ==>(NEG) fins=false 58
79. breathes=true 80 ==>(NEG) fins=false 76
80. backbone=true breathes=true tail=true 60 ==>(NEG) fins=false 57
81. fins=false tail=true 59 ==>(NEG) venomous=false 56
82. fins=false tail=true 59 ==>(NEG) backbone=true venomous=false 56
83. breathes=true venomous=false tail=true 59 ==>(NEG) fins=false 56
84. fins=false tail=true 59 ==>(NEG) breathes=true venomous=false 56
85. backbone=true breathes=true venomous=false tail=true 59 ==>(NEG) fins=false 56
86. breathes=true venomous=false tail=true 59 ==>(NEG) backbone=true fins=false 56
87. fins=false tail=true 59 ==>(NEG) backbone=true breathes=true venomous=false 56
88. airborne=false 77 ==>(NEG) feathers=false 73
89. tail=true 75 ==>(NEG) venomous=false 71
90. tail=true 75 ==>(NEG) backbone=true venomous=false 71
91. breathes=true venomous=false 75 ==>(NEG) fins=false 71
92. airborne=false venomous=false 71 ==>(NEG) feathers=false 67
93. backbone=true domestic=false 71 ==>(NEG) venomous=false 67
94. backbone=true breathes=true 69 ==>(NEG) fins=false 65
95. airborne=false domestic=false 68 ==>(NEG) feathers=false 64
96. breathes=true domestic=false 68 ==>(NEG) venomous=false 64
97. breathes=true domestic=false 68 ==>(NEG) fins=false 64
98. backbone=true breathes=true venomous=false 67 ==>(NEG) fins=false 63
99. aquatic=false 65 ==>(NEG) venomous=false 61
100. airborne=false backbone=true 65 ==>(NEG) feathers=false 61

=== Evaluation ===

Elapsed time: 0.04s

POSITIVE COHERENT BASED RULES:
======

feathers=false backbone=true 63 ==> airborne=false 61
feathers=false backbone=true 63 ==> toothed=true 61
backbone=true tail=true 74 ==> venomous=false 71
backbone=true fins=false 66 ==> venomous=false 63
backbone=true fins=false 66 ==> breathes=true venomous=false 63
backbone=true tail=true domestic=false 64 ==> venomous=false 61
backbone=true 83 ==> venomous=false 79
breathes=true 80 ==> fins=false 76
airborne=false 77 ==> feathers=false 73
tail=true 75 ==> venomous=false 71
tail=true 75 ==> backbone=true venomous=false 71
breathes=true venomous=false 75 ==> fins=false 71
airborne=false venomous=false 71 ==> feathers=false 67
backbone=true domestic=false 71 ==> venomous=false 67
backbone=true breathes=true 69 ==> fins=false 65
airborne=false domestic=false 68 ==> feathers=false 64
breathes=true domestic=false 68 ==> venomous=false 64
breathes=true domestic=false 68 ==> fins=false 64
backbone=true breathes=true venomous=false 67 ==> fins=false 63

=== Evaluation ===

*Elapsed time: 0.02s*

**6. CONCLUSION AND FUTURE WORK**

We conclude from our design of a threshold free association rule mining technique that a minimum support threshold . We used mapping to logical equivalences according to propositional logic to discover all interesting association rules without loss. These association rules include item sets that are frequently and infrequently observed in a set of transaction records. In addition to a complete set of rules being considered, these association rules can also be reasoned as logical implications because they inherit propositional logic properties. Having considered infrequent items, as well as being implicational, these newly discovered association rules are distinguished from typical association rules. The framework is developed based on implication of propositional logic via Negative and positive association algorithm. The experiments show that our approach is able to identify meaningful association rules within an acceptable execution time. This framework develop a new algorithm based on coherent rules so that users can mine the items without domain knowledge and it can mine the items efficiently when compared to association rules. Implication of propositional logic is a good alternative on the definition on association. Rules based on this definition may be searched and discovered within feasible time.

In future this work is extended to implement current framework with the classification based associative algorithms in order to get effective classification based rules.In future ecommerce based application are used to check the product relationship for customer based analysis.

**REFERENCES:**

[1] Jaiwei Han and Micheline Kamber, "Data Mining Concepts and Techniques", Second Edition , Morgan Kaufmann Publishers.

[2] Feng Yucai, "Association Rules Incremental Updating Algorithm", Journal of Software, Sept., 1998.

[3] Association Rule Learning – Wikipedia, the free encyclopedia. [4] Lei Guoping, Dai Minlu, Tan Zefu and Wang Yan, " The Research of CMMB Wireless Network Analysis Based on Data Mining Association Rules", IEEE conference paper – project supported by the Science and Technology Research Project of Chongqing municipal education commision under contract no KJ101114 and KJ 111103, 2011

[5] Jayalakshmi.S, Dr k. Nageswara Rao, "Mining Association rules for Large Transactions using New Support and Confidence Measures", Journal of Theoretical and applied Information Technology, 2005. [6] Hou Sizu, Zhang Xianfei, " Alarms Association Rules Based on Sequential Pattern Mining Algorithm", Fifth IEEE International Conference on fuzzy Systems and Knowledge Discovery, 2008.

[6] Srikant, R. & Agrawal, R., "Mining quantitative association rules in large relational tables", *SIGMOD Rec.,* ACM, 1996, 25, 1-12.

[7] Castelo, R.; Feelders, A. & Siebes, A., "MAMBO: Discovering Association Rules Based on Conditional Independencies", *Lecture Notes in Computer Science*, Springer Berlin / Heidelberg, 2001, 2189, 289-298.

[8] Lin, W.; Tseng, M. & Su, J., "A Confidence-Lift Support Specification for Interesting Associations Mining", *PAKDD '02: Proceedings of the 6th Pacific-Asia Conference on Advances in Knowledge Discovery and Data Mining*, Springer-Verlag, 2002, 148-158.

[9] Brin, S.; Motwani, R. & Silverstein, C., "Beyond market baskets: generalizing association rules to correlations", *SIGMOD Rec.*, ACM, 1997, 26, 265-276.

[10] Zhou, L. & Yau, S., "Efficient association rule mining among both frequent and infrequent items", *Comput. Math. Appl.*, Pergamon Press, Inc., 2007, 54, 737-749.

[11] Wu, X.; Zhang, C. & Zhang, S., "Mining Both Positive and Negative Association Rules", *ICML'02*, 2002, pp. 658-665.